\documentclass[aps,prb,twocolumn, showpacs]{revtex4}

\usepackage{epsfig}
\usepackage{graphicx}

\def\be{\begin{equation}}
\def\ee{\end{equation}}
\def\bea{\begin{eqnarray}}
\def\eea{\end{eqnarray}}

\begin{document}
%\draft

\title{Quantum nanostructures in strongly spin-orbit coupled two-dimensional systems}
\author{Emmanuel I. Rashba}
\affiliation{Department of Physics, Harvard University, Cambridge, Massachusetts 02138, USA}  
\date{\today}
\begin{abstract}
Recent progress in experimental studies of low-dimensional systems with strong spin-orbit coupling poses a question on the effect of this coupling on the energy spectrum of electrons in semiconductor nanostructures. It is shown in the paper that this effect is profound in the strong coupling limit. In circular quantum dots a soft mode develops, in strongly elongated dots electron spin becomes protected from the effects of the environment, and the lower branch of the energy spectrum of quantum wires becomes nearly flat in a wide region of the momentum space. 
\end{abstract}


\maketitle

\narrowtext

\section{Introduction}

Quantum dots are basic elements of the systems aimed in semiconductor based quantum computation based on qubits encoded in electron spins.\cite{LDV,KouMar98,Awsch02} Electrical operation of such qubits is possible through the Heisenberg (exchange) mechanism,\cite{DiV2000,Laird2010,Stu2012} or spin-orbit coupling, or both.\cite{Petta2005} Effective spin-orbit coupling in quantum dots originates from two different mechanisms. First mechanism is based on the celebrated Thomas term coupling electron momentum $\bf p$ to Pauli matrices $\mbox{\boldmath$\sigma$}$. In two-dimensional (2D) systems it is usually represented through two invariants linear in the quasimomentum $\bf k$, known as the Rashba and Dresselhaus terms, reflecting the structure, bulk, and interface induced asymmetry.\cite{EDSR} Second mechanism couples  $\mbox{\boldmath$\sigma$}$ to the electron coordinate $\bf r$ through the inhomogeneous external, exchange, or hyperfine magnetic field.\cite{PR1965,Tokura06} Electric dipole spin resonances (EDSR)\cite{EDSR,R60,Golovach06,Rashba08}  driven by these mechanisms 
have been observed experimentally\cite{KMDGLA03,Nowack07,Laird07,Tarucha08,NP2012,Vandersypen} in different semiconductor nanostructures. Electric fields generated by precessing spins are an inverse effect but nediated by the same mechanisms.\cite{LR2003}

Effect of spin-orbit coupling on the spin dynamics inside a dot can be quantified by the ratio of the dot size $d$ to the spin-orbit length $\ell_{SO}$. For GaAs, spin-orbit coupling is relatively weak, with $\ell_{SO}\sim 35~ \mu$m, hence, with $d\sim 100$ nm, $d/\ell_{SO}\sim10^{-3}\ll1$.\cite{Nowack07} Currently, interest is shifting in the direction of semiconductor dots from the materials with much stronger spin-orbit coupling such as InAs and InSb; for them $\ell_{SO}\sim300 $ nm.\cite{NP2012} For all materials with $d/\ell_{SO}\alt1$, spin-orbit coupling influences mostly the spin dynamics and has only a minor effect on the shape of the electron cloud inside the dot. 

However, recently new classes of systems emerged, based on the chemical elements with high atomic numbers, where spin-orbit coupling is even much stronger, up to the atomic scale. They are considered as new potential materials for spintronics,\cite{Zutic} in particular, for the Datta-Das spin transistor.\cite{Datta1990} If to estimate spin-orbit energy in the Hamiltonian as $\alpha_{SO}k$, in such systems $\alpha_{SO}\agt 1$ eV~\AA. These are noncentrosymmetric bulk BiTeI with $\alpha_{SO}=3.85$ eV \AA,\cite{BiTeI} Te-terminated surface of BiTeCl with $\alpha=1.78$ eV \AA\, and the minimum of the surface electron band well inside the bulk gap,\cite{BiTeCl} HgTe quantum wells with $k$-dependent $\alpha$ reaching the value $\alpha\alt1$ eV \AA,\cite{Gui} and interfaces between insulating perovskite oxides LaAlO$_3$/SrTiO$_3$ with $\alpha_{SO}\alt5\times10^{-2}$ eV \AA,\cite{Ohtomo,Caviglia2010,Loder} to name only a few (topological materials are outside the scope of the paper). For nanocrystals of high $\alpha_{SO}$ compounds, and quantum dots in them, an opposite inequality, $d/\ell_{SO}\agt1$, should be fulfilled, with the charge distribution strongly influenced by spin-orbit coupling; for quantum dots (or nanocrystals) of BiTeI and BiTeCl, one expects $d/\ell_{SO}\gg1$. The focus of the present paper is on the electronic structure of strongly spin-orbit coupled 2D nanostructures, but still in the range of the applicability of the envelope function approximation.

\section{Parabolically confined dots}
\label{sec:parab} 

A model Hamiltonian for a lateral quantum dot with parabolic confinement is
\be
{\hat H}=\frac{\hbar^2{\bf k}^2}{2m}+\alpha_{SO}(\mbox{\boldmath$\sigma$}\times{\bf k})_z+\frac{m\omega_0^2}{2}{\bf r}^2
\label{eq1}
\ee
It is convenient to introduce dimensionless units using $\hbar\omega_0$ as a unit of energy, $E=\hbar\omega_0\epsilon$, and $\ell_0=\sqrt{\hbar/m\omega_0}$ as a unit of length. Then, with ${\bf r}=\ell_0{\mbox{\boldmath$\rho$}}$, ${\bf k}=\mbox{\boldmath$\kappa$}/\ell_0$, $\ell_{SO}=\hbar^2/m\alpha_{SO}$, and $\alpha={\ell_0}/{\ell_{SO}}$, the Hamiltonian reduces to
\be
{\hat h}=\frac{\mbox{\boldmath$\kappa$}^2}{2}+\alpha(\mbox{\boldmath$\sigma$}\times\mbox{\boldmath$\kappa$})_z+\frac{\mbox{\boldmath$\rho$}^2}{2}.
\label{eq2}
\ee
The energy spectrum of a free particle [last term in Eq.~(\ref{eq2}) omitted], consists of two branches $\epsilon_0(\kappa)=\kappa^2/2-\alpha\kappa$ with a minimum $\epsilon_\alpha=-\alpha^2/2$ achieved at a circle of the radius of $\kappa_\alpha=\alpha$ in the momentum space. For finding the energy spectrum of a quantum dot near its bottom in the strong spin-orbit coupling limit, $\alpha\gg1$, it is convenient to solve the eigenvalue problem for a two-component spinor $\psi=\left(\begin{array}{c}u\\v\end{array}\right)$ in the momentum representation\cite{Weyl}
\bea
\left[-\frac{1}{2\kappa}\partial_\kappa(\kappa\partial_\kappa)-\frac{1}{\kappa^2}\partial^2_\phi+\frac{\kappa^2}{2}-\epsilon\right]u+i\alpha\kappa_-v&=&0,\nonumber\\
\left[-\frac{1}{2\kappa}\partial_\kappa(\kappa\partial_\kappa)-\frac{1}{\kappa^2}\partial^2_\phi+\frac{\kappa^2}{2}-\epsilon\right]v-i\alpha\kappa_+u&=&0,
\label{eq3}
\eea
with $\kappa_\pm=\kappa e^{\pm i\phi}$. Therefore, $u(\rho,\phi)=u(\rho)e^{i(j-1/2)\phi}$, $v(\rho,\phi)=v(\rho)e^{i(j+1/2)\phi}$, with half-integer $j\geq 1/2$; levels with $j<0$ possess the same energy due to the Kramers symmetry. Equations for the radial parts $(u(\rho),v(\rho))$ are simplified when rewritten for their linear combinations $w_\pm=u\pm iv$ 
\bea
\bigg[-\frac{1}{2\kappa}\partial_\kappa(\kappa\partial_\kappa)+\frac{j^2+1/4}{\kappa^2}+\frac{\kappa^2}{2}&+&\alpha\kappa-\epsilon\bigg]w_+\nonumber\\
&-&\frac{j}{\kappa^2}w_-=0,\nonumber\\
\bigg[-\frac{1}{2\kappa}\partial_\kappa(\kappa\partial_\kappa)+\frac{j^2+1/4}{\kappa^2}+\frac{\kappa^2}{2}&-&\alpha\kappa-\epsilon\bigg]w_-\nonumber\\
&-&\frac{j}{\kappa^2}w_+=0.
\label{eq4}
\eea

It is seen from Eq.~(\ref{eq3}) that near the spectrum bottom $\vert w_-\vert\gg\vert w_+\vert$, and $w_+\approx(j/2\alpha^4)w_-$. Because in what follows the expansion will go in the parameter $j/\alpha\ll1$, Eqs.~(\ref{eq4}) decouple. In the same spectrum region, the equation for $w_-$ reduces to
\be
\left[\left(-\frac{1}{2}\partial^2_\kappa+\frac{\kappa^2}{2}-\alpha\kappa-\epsilon\right)-\frac{1}{2\alpha}\partial_\kappa+\frac{j^2+1/4}{\alpha^2}\right]w_-=0.
\label{eq5}
\ee
First term, enclosed in parentheses, is a harmonic oscillator related to the radial motion with the spectrum
$\epsilon_n=(n+1/2)-\alpha^2/2,\,\,n\geq0.$
Second term produces, in 2nd order of the perturbation theory, an $n$-independent level shift $\Delta\epsilon=-1/8\alpha^2$; it plays no significant role. Third term describes a soft mode related to the angular degree if freedom with a typical level spacing of $\approx2j/\alpha^2$; first gap, separating the ground and first excited states, equals $2/\alpha^2\ll1$. Finally
\be
E_n-E_\alpha=\hbar\omega_0\left[(n+1/2)-\frac{j^2+1/8}{\alpha^2}\right]
\label{eq6}
\ee
with $E_\alpha=-\hbar\omega_0\alpha^2/2$, $n\geq0$, and $j\geq1/2$. Small separation between the energy levels with the same $n$ but different values of $j$ hampers trapping of a single electron into a well defined quantum state.

For comparison, in absence of spin-orbit coupling, $\alpha=0$, the Hamiltonian of Eq.~(\ref{eq2}) describes a two-dimensional harmonic oscillator. Its ground state is well protected by gaps of which the smallest one equals to $\hbar\omega_0$.

The above results prove that, in axially symmetric systems with a strong spin-orbit coupling, $\alpha\gg1$, confinement in quantum dots is hindered by the motion along the angular degree of freedom, and achieving a well isolated ground state requires freezing angular dynamics by breaking the axial symmetry. Angular dynamics is completely suppressed in a one-dimensional setup with a Hamiltonian
\be
{\hat h}=\frac{\kappa^2_\eta}{2}+\alpha\sigma_1\kappa_\eta+\frac{\eta^2}{2},\,\,\eta=y/\ell_0,
\label{eq7}
\ee
that follows from Eq.~(\ref{eq2}) by suppressing first Cartesian coordinate in $\mbox{\boldmath$\rho$}=(\xi,\eta)$. It possesses the harmonic oscillator energy spectrum $\epsilon_n=(n+1/2)$ with the ground state eigenfunctions
\be
\psi_1(\eta)=\frac{1}{\pi^{1/4}}\left(\begin{array}{c}\cos\alpha\eta\\-i\sin\alpha\eta\end{array}\right)e^{-\eta^2/2},\,\,\psi_2=\sigma_2\psi_1,
\label{eq8}
\ee
forming a Kramers doublet; $(\sigma_1, \sigma_2)$ are two first Pauli matrices. Hence, the energy spectrum is well gapped (with the gap $\Delta E=\hbar\omega_0$), the electron density is a smooth function of the coordinate and is trapped near the center of the parabolic well at the scale $\Delta y\sim\ell_0$, while the mutual phase of the spinor components oscillates with a frequency controlled by the spin-orbit coupling constant $\alpha$. Fourier images of $(\psi_1,\psi_2)$ are peaked at $\kappa_\eta=\pm i\alpha$.

However, as is explained in Sec.~\ref{sec:wire} below, the Hamiltonian of Eq.~(\ref{eq7}) is not a proper model Hamiltonian for a quasi-one-dimensional quantum dot in a strongly spin-orbit coupled material described by the Hamiltonian of Eq.~(\ref{eq2}). In particular, the Fourier images of wave functions are peaked at $\kappa_\eta=\pm\alpha$, hence, they imply a state that is a double dot in the momentum space. The origin of this striking difference comes from the fact that two first terms of the Hamiltonian $\hat h$ of Eq.~(\ref{eq7}) with $\alpha={\rm const}$ do not describe the dispersion of the lower branches of the energy spectrum of a quantum wire that are flat in a wide region of the momentum space $\vert\kappa_y\vert\alt\alpha$, see Fig.~1. Unusual properties of quasi-one-dimensional quantum dots in strongly spin-orbit coupled materials are discussed at the end of Sec.~\ref{sec:wire}.

\section{Hard-wall confined dots}
\label{sec:hard}

Hard-wall confinement provides an independent insight onto the same problem. It has been already investigated by Bulgakov and Sadreev\cite{BS2001} and Tsitsishvili et al.\cite{TLG2004} Therefore, technical details will be omitted in what follows, and we focus on the large $\alpha_{SO}$ limit; the Hamiltonian is the same as in Sec.~\ref{sec:wire} below. After choosing the radius $R$ of the confinement region as the unit of the length,  introducing dimensionless quantities $\rho=r/R$, $\epsilon=E/(\hbar^2/mR^2)$, $\alpha=R/\ell_{SO}$, and separating the angular variable $\phi$ similarly to Sec.~\ref{sec:parab}, equations for the eigenspinor components $(u,v)$ are
\bea
[\rho^2\partial^2_\rho+\rho\partial_\rho&-&(j-1/2)^2+2\epsilon\rho^2]u\nonumber\\
&-&2\alpha\rho^2[\partial_\rho+\frac{j+1/2}{\rho}]v=0,\nonumber
\eea
\bea
[\rho^2\partial^2_\rho+\rho\partial_\rho&-&(j+1/2)^2+2\epsilon\rho^2]v\nonumber\\
&+&2\alpha\rho^2[\partial_\rho-\frac{j-1/2}{\rho}]u=0.
\label{eq9}
\eea
Due to the recurrence relations\cite{Erdelyi} 
\be
[\partial_\rho\pm(j\pm1/2)/\rho]J_{j\pm1/2}(\kappa\rho)=\pm\kappa J_{j\mp1/2}(\kappa\rho),
\label{eq10}
\ee
solutions of Eqs.~(\ref{eq9}) are Bessel functions $u(\rho)=aJ_{j-1/2}(\kappa\rho)$, $v(\rho)=bJ_{j+1/2}(\kappa\rho)$, with half-integer values of the angular momentum $j\geq1/2$. The hard-wall boundary condition $u(1)=v(1)=0$ results in the following equation\cite{BS2001,TLG2004} for the energy eigenvalues $\epsilon$
\be
J_{j-1/2}(\kappa_+)J_{j+1/2}(\kappa_-)=J_{j-1/2}(\kappa_-)J_{j+1/2}(\kappa_+),
\label{eq11}
\ee
with
\be
\kappa_\pm=\alpha\pm\sqrt{\alpha^2+2\epsilon}\,.
\label{eq12}
\ee
Equation (\ref{eq11}) was derived\cite{BS2001,TLG2004} under the condition of the existence of two spectrum branches, therefore, its trivial solution $\kappa_+=\kappa_-$, corresponding to the free-particle spectrum bottom, should be omitted.

Below the conical (Dirac) point of the spectrum, for $-\alpha^2/2<\epsilon<0$, both $\kappa_\pm>0$. Near the spectrum bottom, $\kappa_\pm\approx\alpha\gg1$, energy spectrum can be found from the $\kappa\gg1$ asymptotic expansion of Bessel functions\cite{Erdelyi}
\bea
J_\nu(\kappa)&\approx&\sqrt{\frac{2}{\pi\kappa}}\bigg[\cos\left(\kappa-\frac{\pi\nu}{2}-\frac{\pi}{4}\right)\nonumber\\
&-&\frac{\nu^2-1/4}{2\kappa}\sin\left(\kappa-\frac{\pi\nu}{2}-\frac{\pi}{4}\right)\bigg].
\label{eq13}
\eea
In the leading order in $1/\kappa\ll1$, with $\epsilon$ close to the spectrum bottom $\epsilon_\alpha=-\alpha^2/2$, when $\epsilon-\epsilon_\alpha\ll\alpha$ and $j \ll\alpha$, Eq.~(\ref{eq11}) reduces to
\bea
j\sqrt{2(\epsilon-\epsilon_\alpha)}&\{&\cos(2\alpha-j\pi)+j\cos[2\sqrt{2(\epsilon-\epsilon_\alpha)}]\}\nonumber\\
&=&\alpha^2\sin[2\sqrt{2(\epsilon-\epsilon_\alpha)}].
\label{eq14}
\eea 

In the leading order in $1/\alpha\ll1$, solutions of Eq.~(\ref{eq14}) are $2\sqrt{2(\epsilon-\epsilon_\alpha)}\approx\pi n$, with $n\geq1$. In the next order, for $\sqrt{jn}\ll\alpha$, the energy spectrum near its bottom $E_\alpha=-\hbar^2/(2m\ell^2_{SO})$, expressed in dimensional units, is
\be
E_{n,j}-E_\alpha\approx\frac{\hbar^2\pi^2n^2}{8mR^2}\left\{1+\frac{j}{\alpha^2}[\cos(2\alpha-j\pi)+(-)^nj]\right\},
\label{eq15}
\ee
with integer $n\geq1$ and half-integer $j\geq1/2$. It is seen that there are two quantum numbers. 

Integers $n$ quantize the radial motion similarly to a particle confined in a hard-wall box in absence of spin-orbit coupling. This coupling does not change the scale of the level spacings of the radial motion that is set by the radius $R$ of the confinement region.  It is seen from the form $J_{j\pm1/2}(\kappa\rho)$ of the solutions of Eq.~(\ref{eq9}) that eigenfunctions are oscillating functions of $\rho$.

Half-integers $j$ describe angular motion. Remarkably, second term in the braces including the angular quantum number $j$ shows oscillatory dependence on $\alpha$. Similarly to Eq.~(\ref{eq6}), the angular mode is soft with the level separation small in $1/\alpha^2$.

For comparison, in absence of spin-orbit coupling, $\alpha=0$, the energy spectrum is described by a Bessel equation similar to Eq.~(\ref{eq9}) but with integer angular momenta $j$. The ground state is protected by gaps of which the smallest one is approximately equal to $4.45\hbar^2/mR^2$.

Equation (\ref{eq15}) is valid only near the spectrum bottom, $\epsilon_{n,j}-\epsilon_\alpha\ll\epsilon_\alpha$. Numerical analysis of Eq.~(\ref{eq11}) shows that with increasing $j$ its roots move up, and the total number of negative roots decreases. Finally, at $j\sim\alpha$, there remains a single negative root $\epsilon_{1,j}<0$ that with increasing $j$ asymptotically approaches zero, $\epsilon_{1,j}\rightarrow0$.

\section{Quantum wires}
\label{sec:wire}

Interest in the energy spectrum of single- and multi-channel spin-orbit coupled quantum wires increased recently after they have been proposed as a platform for solid state realization of Majorana fermions.\cite{Lutchyn,Oreg,Akhmerov,Potter,Alicea}

For a hard-wall confined quantum wire of a width $Y$, $-Y/2\leq y\leq Y/2$, it is convenient to choose $Y$ as a unit of length and $\hbar^2/mY^2$ as a unit of energy, $E=(\hbar^2/mY^2)\epsilon$. Then, with dimensionless $\alpha$ defined as $\alpha=Y/\ell_{SO}$, the Hamiltonian is
\be
{\hat h}=\frac{{\bf k}^2}{2}+\alpha(\mbox{\boldmath$\sigma$}\times{\bf k})_z.
\label{eq16}
\ee
In Eq.~(\ref{eq16}), $\bf r$ and $\bf k$ are dimensionless coordinate and momentum, respectively.

For energies below the conical (Dirac) point, $\epsilon<0$, eigenfunctions and energy of a free particle are
\be
\psi_{\bf k}({\bf r})=\frac{1}{\sqrt{2}}\left(\begin{array}{c}1\\i(k_x+ik_y)/k\end{array}\right)e^{i({\bf k}\cdot{\bf r})},\,\epsilon_0(k)=\frac{k^2}{2}-\alpha k.
\label{eq17}
\ee
It is convenient to define momenta
\be
k_\pm(\epsilon)=\alpha\pm\sqrt{\alpha^2+2\epsilon}
\label{eq18}
\ee
at the external (internal) arcs of the energy spectrum. Then, for $k_x<k_-$, eigenfunctions of confined states can be chosen as linear combinations of four functions $\psi_{\bf k}({\bf r})$ with 
\be
k_y^\pm(\epsilon, k_x)=\sqrt{k_\pm^2(\epsilon)-k_x^2}
\label{eq19}
\ee
 and $-k_y^\pm(\epsilon, k_x)$, obeying the boundary conditions $\psi_{\bf k}(x,\pm1/2)=0$. After some algebra, the eigenvalue problem reduces to
\be
(k_x^2+2\epsilon)\sin k_y^-\sin k_y^+
+k_y^-k_y^+(1-\cos k_y^-\cos k_y^+)=0.
\label{eq20}
\ee

Equation (\ref{eq20}) can be solved exactly for $k_x=0$. Indeed, in this case $k_y^-k_y^+=2\vert\epsilon\vert$ and, for $\epsilon<0$, Eq.~(\ref{eq20}) reduces to $\cos(k_+-k_-)=1$. Its solutions are
\be
k_+(\epsilon)-k_-(\epsilon)=2\pi n,\,\,n\geq1.
\label{eq21}
\ee
The value $n=0$ is excluded because it would result in   
$k_+(\epsilon)=k_-(\epsilon)$ and decreasing the number of linearly independent basis functions.  Squaring Eq.~(\ref{eq21}) leads to
\be
\epsilon_n(k_x=0)=-\frac{\alpha^2}{2}+\frac{\pi^2n^2}{2},\,\,\,\, n\geq1.
\label{eq22}
\ee
Therefore, for $k_x=0$ the energy spectrum reduces to the spectrum of the textbook quantum box problem, and spin-orbit coupling manifests itself only through lowering the spectrum bottom to $\epsilon_\alpha=-\alpha^2/2$. Energy levels of Eq.~(\ref{eq22}) are Kramers degenerate and split with increasing $k_x$ as seen in Fig.~1. 

The spectrum described by Eq.~(\ref{eq20}) extends until $k_x=k_-(\epsilon)$, which corresponds to $\epsilon=k_x^2/2-\alpha k_x$. This boundary is shown by a heavy line in Fig.~1 and describes a free particle moving in the $x$ direction. To the right from this line, propagating waves $\exp(\pm ik_y^-y)$ change to evanescent waves $\exp(\pm\kappa_y y)$ with $\kappa_y=\sqrt{k_x^2-k_-^2(\epsilon)}$, and Eq.~(\ref{eq20}) should be substituted with
\bea
(k_x^2+2\epsilon)\sinh \kappa_y\sin k_y^+
+\kappa_y k_y^+(1-\cosh \kappa_y\cos k_y^+)=0.\nonumber\\
\label{eq23}
\eea

Energy spectrum for $\alpha=10$ is shown in Fig.~1. At $k_x=0$, there are three Kramers degenerate energy levels with energies $\epsilon<0$ that split with increasing momentum $k_x$. For the $n=1$ levels, this splitting is very small until $k_x\approx8$, and there are two level intersections, near $k_x=3.0$ and $k_x=7.4$. These are real intersections rather than avoided crossings because the Hamiltonian of Eq.~(\ref{eq16}) possesses a $xz$ symmetry plane and two $n=1$ branches have opposite parity with respect to the reflections in this plane. The splitting only increases for $\alpha\agt8$ when the momentum approaches the value $k_x=\alpha$ where the free-particle dispersion law $\epsilon_0(k_x)=k_x^2/2-\alpha k_x$ reaches its minimum $\epsilon_\alpha=-\alpha^2/2$. At $k_x=\alpha$, deviation of the lower branch from $\epsilon_\alpha$ is only $\epsilon(k_x=\alpha)-\epsilon_\alpha\approx105/2\alpha^2$. The absolute minimum of the lower branch $\epsilon_{\rm min}\approx\epsilon_\alpha + 0.411$ is achieved slightly to the left from this point, at $k_x\approx9.442$.

The left arc of the free-particle dispersion curve $\epsilon_0(k_x)$ cuts the plane into two parts. To the left from it quantum states are described by solutions of Eq.~(\ref{eq20}) and to the right by solutions of Eq.~(\ref{eq23}).

Flat shape of both $n=1$ curves in Fig.~1, together with the position of the  spectrum minimum near $k_x=\alpha\gg1$, suggests unusual properties of the quantum dots prepared from large-$\alpha$ quantum wires. We discuss them below at a qualitative level.

The curvature of the lower $n=1$ branch near its minimum is close to that of the $\epsilon_0(k_x)$ curve, i.e., close to unity. For localizing a particle close to the minimum, the width of the quantum state in the $k_x$ space should be $\Delta k_x\ll\pi$ as it follows from Eq.~(\ref{eq22}). Hence, according to the uncertainty principle, the length of the dot $X$ (in dimensional units) should satisfy the condition
\be
X\gg Y/\pi.
\label{eq24}
\ee
If this condition is satisfied with sufficient accuracy, the ground state of the dot possesses highly unusual properties. This is a Kramers doublet comprising the states centered at $k_x=\pm\alpha$ with spins polarized parallel to $\hat{\bf y}$ in opposite directions; it resembles a double dot but in the momentum rather than in the real space. Transitions between these states are only possible through tunneling in the $k_x$ space under a barrier of the hight of nearly $\pi^2/2$ and the width of about $2\alpha$. Therefore, we encounter a unique situation when strong spin-orbit coupling protects electron spin against decoherence. Applying a magnetic field ${\bf B}\parallel{\hat{\bf y}}$ splits the ground state into a Zeeman doublet. Remarkably, the intensities of both the electron spin resonance and EDSR transitions between the components of the doublet should be strongly suppressed because the tunneling process is involved.\cite{conjec} Therefore, in this unique regime electron spin becomes well protected against low-frequency electric and magnetic noise by strong spin-orbit coupling.

With decreasing $X$, wave function spreads fast in the $k_x$ space, mostly in the direction of small $k_x$ values, involving the upper $n=1$ branch, and than also next branches of the energy spectrum. As $X$ approaches $Y$, the shape of the dot becomes close to circular, and a soft mode discussed in Secs.~\ref{sec:parab} and \ref{sec:hard} becomes pronounced.

In conclusion, with the increasing strength of spin-orbit coupling, when the spin-orbit length becomes shorter than the nanostructure size, this coupling influences heavily the energy spectrum and electron density distribution. In the strong spin-orbit coupling limit, circular quantum dots acquire a soft mode, dispersion of the lower mode of quantum wires becomes flat in a wide region of the momentum space, and electron spin becomes protected from the effects of the environment in strongly elongated quantum dots.

Funding from the NSF Materials World Network and the Intelligence Advanced Research Project Activity (IARPA) through the Army Research Office is gratefully acknowledged.


\begin{thebibliography}{99}
\bibitem{LDV} D. Loss and D. P. DiVincenzo, Phys. Rev. A {\bf 57}, 120 (1998).
\bibitem{KouMar98} L. P. Kouwenhoven and C. Marcus, Physics World {\bf 11} (6), 35 (1998).
\bibitem{Awsch02} D. D. Awschalom, D. Loss, and N. Samarth, {\it Semiconductor Spintronics and Quantum Computation}, 2002 (Springer, Berlin).
\bibitem{DiV2000} D. P. DiVincenzo, D. Bacon, J. Kempe, G. Burkard, and K. B. Whaley, Nature {\bf 408}, 339 (2000).
\bibitem{Laird2010} E. A. Laird, J. M. Taylor, D. P. DiVincenzo, C. M. Marcus, M. P. Hanson, and A. C. Gossard, Phys. Rev. B {\bf 82}, 075403 (2010).
\bibitem{Stu2012} S. A. Studenikin, G. C. Aers, G. Granger, L. Gaudreau, A. Kam, P. Zawadzki, Z. P. Wasilewski, and A. S. Sachrajda, Phys. Rev. Lett. {\bf 108}, 226802 (2012).
\bibitem{Petta2005} J. R. Petta, A. C. Johnson, J. M. Taylor, E. A. Laird, A. Yacoby, M. D. Lukin, C. M. Marcus, M. P. Hanson, and A. C. Gossard, Science {\bf 309}, 2180 (2005).
\bibitem{RJSup2005} E. I. Rashba, J. Supercond. {\bf 18}, 137 (2005).
\bibitem{EDSR} E. I. Rashba and V. I. Sheka: {\it Electric Dipole Spin Resonances}, in: Landau Level Spectroscopy, 1991 (North-Holland, Amsterdam), p. 131.
\bibitem{PR1965} S. I. Pekar and E. I. Rashba, Zh. Eksp. Teor. Fiz. {\bf 47}, 1927 (1964) [Sov. Phys. - JETP {\bf 20}, 1295 (1965)].
\bibitem{Tokura06} Y. Tokura, W. G. van der Wiel, T. Obata, and S. Tarucha, Phys. Rev. Lett. {\bf 96}, 047202 (2006).
\bibitem{R60} E. I. Rashba, Fiz. Tverd. Tela {\bf 2}, 1224 (1960) [Sov. Phys. Solid State {\bf 2}, 1109 (1960)].
\bibitem{Golovach06} V. N. Golovach, M. Borhani, and D. Loss, Phys. Rev. B {\bf 74}, 165319 (2006).
\bibitem{Rashba08} E. I. Rashba, Phys. Rev. B {\bf 78}, 195302 (2008).
\bibitem{KMDGLA03} Y. Kato, R. C. Myers, D. C. Driscoll, A. C. Gossard, J. Levy, and D. D. Awschalom, Science {\bf 299}, 1201 (2003).
\bibitem{Nowack07} K. C. Nowack, F. H. L. Koppens, Y. V. Nazarov, and L. M. K. Vandersypen, Science  {\bf 318},  1430 (2007).
\bibitem{Laird07} E. A. Laird, C. Barthel, E. I. Rashba, C. M. Marcus, M. P. Hanson, and A. C. Gossard, Phys. Rev. Lett. {\bf 99}, 246601 (2007).
\bibitem{Tarucha08} M. Pioro-Ladri\`{e}re, T. Obata, Y. Tokura, Y.-S. Shin, T. Kubo, K. Yoshida, T. Taniyama, and S. Tarucha, {\it Nature Physics} {\bf 4}, 776 (2008).
\bibitem{NP2012} S. Nadj-Perge, V. S. Pribiag, J.W. G. van den Berg, K. Zuo, S. R. Plissard, E. P. A. M. Bakkers, S. M. Frolov, and L. P. Kouwenhoven, Phys. Rev. Lett. {\bf 108}, 166801 (2012).
\bibitem{Vandersypen} M. Shafiei, K. C. Nowack, C. Reichl, W. Wegscheider, L. M. K. Vandersypen,  arXiv:1207.3331. 
\bibitem{LR2003} L. S. Levitov and E. I. Rashba, Phys. Rev. B {\bf 67}, 115324 (2003).
\bibitem{Zutic} I. \v{Z}uti\'{c}, J. Fabian, and S. Das Sarma, Rev. Mod. Phys. {\bf 76}, 323 (2004).
\bibitem{Datta1990} S. Datta and B. Das,  Appl. Phys. Lett. {\bf 56}, 665 (1990).
\bibitem{BiTeI} K. Ishizaka, M. S. Bahramy, H. Murakawa, M. Sakano, T. Shimojima, T. Sonobe, K. Koizumi, S. Shin, H. Miyahara, A. Kimura, K. Miyamoto, T. Okuda, H. Namatame, M. Taniguchi,
R. Arita, N. Nagaosa, K. Kobayashi, Y. Murakami, R. Kumai, Y. Kaneko, Y. Onose, 
and Y. Tokura, Nature Mater. {\bf 10}, 521 (2012).
\bibitem{BiTeCl} S.V. Eremeev, I. A. Nechaev, Yu. M. Koroteev, P. M. Echenique, and E. V. Chulkov, Phys. Rev. Lett. {\bf 108}, 246802 (2012).
\bibitem{Gui} Y. S. Gui, C. R. Becker, N. Dai, J. Liu, J. Qiu, E. G. Novik, M. Sch\"{a}fer, X. Z. Shu, J. H. Chu, H. Buhmann, and L. W. Molenkamp, Phys. Rev. B {\bf 70}, 115328 (2004).
\bibitem{Ohtomo} A. Ohtomo and H. Y. Hwang, Nature {\bf 427}, 423 (2004).
\bibitem{Caviglia2010} A. D. Caviglia, M. Gabay, S. Gariglio, N. Reyren, C. Cancellieri, and J.-M. Triscone, Phys. Rev. Lett. {\bf 104}, 126803 (2010).
\bibitem{Loder} F. Loder, A. P. Kampf, and T. Kopp, arXiv:1206:1816.
\bibitem{Weyl} A similar approach has been used by Anderson and Clark for solving the problem of a synthetic three-dimensional harmonic potential with the Weyl spin-orbit coupling.\cite{AC2012}
%\bibitem{AC2012} B. M. Anderson and C. W. Clark, arXiv:1206.0018.
\bibitem{BS2001} E. N. Bulgakov and A. F. Sadreev, JETP Lett. {\bf 73}, 505 (2001).
\bibitem{TLG2004} E. Tsitsishvili, G. S. Lozano, and A. O. Gogolin, Phys. Rev. B {\bf 70}, 115316 (2004). 
\bibitem{Erdelyi} {\it Higher Transcendental Functions}, edited by A. Erd\'{e}lyi, 1953 (Mc Graw-Hill, N.Y.), Chapter 7. 
\bibitem{Lutchyn} R. M. Lutchyn, J. D. Sau, S. Das Sarma, Phys. Rev. Lett.
{\bf 105}, 077001 (2010).
\bibitem{Oreg} Y. Oreg, G, Refael, F. von Oppen, Phys. Rev. Lett. {\bf 105},
177002 (2010).
\bibitem{Potter} A. C. Potter and P. A. Lee, Phys. Rev. B {\bf 83}, 094525 (2011).
\bibitem{Akhmerov} A. R. Akhmerov, J. P. Dahlhaus, F. Hassler, M. Wimmer, and C. W. J. Beenakker, Phys. Rev. Lett. {\bf 106}, 057001 (2011).
\bibitem{Alicea} J. Alicea, Rep. Progr. Phys. {\bf 75}, 076501 (2012).
\bibitem{conjec} Such a behavior suggests a conjecture that the EDSR intensity reaches its maximum for $\ell_{SO}$ comparable to the nanostructure size. It requires additional justification.
\bibitem{AC2012} B. M. Anderson and C. W. Clark, arXiv:1206.0018.

\vspace {1cm}

FIG. 1. (Color online) Energy spectrum of a strongly spin-orbit coupled quantum wire. Energy $\epsilon(k_x)$ plotted {\it vs} momentum $k_x$ for the coupling constant value $\alpha=10$; dimensionless units. Only negative part of the spectrum, below the conical point, is shown. Energy levels at $k_x=0$ are numerated by a quantum number $n\geq1$. Two branches emerging from each $k_x=0$ quantum level are shown by full and dashed lines. Heavy (red) line shows free-particle energy $\epsilon_0(k_x)=k_x^2/2 -\alpha k_x$. Notice week dispersion of both $n=1$ branches.



\end{thebibliography}
\end{document}